\documentclass[column, aps, showpacs, showkeys, nofootinbib, floatfix]{revtex4}

\usepackage{amssymb}
\usepackage{amsmath}
\usepackage{graphicx}
\begin{document}
\title{The influence of quintessence on the motion of a binary system in cosmology}

\author{Fei Yu}
\email{yufei1980@student.dlut.edu.cn}
\author{Molin Liu}
\email{mlliudl@student.dlut.edu.cn}
\author{Yuanxing Gui}
\email{guiyx@dlut.edu.cn}

\affiliation{School of Physics and Optoelectronic Technology, Dalian
University of Technology, Dalian, 116023, P. R. China}

\begin{abstract}
We employ the metric of Schwarzschild space surrounded by
quintessential matter to study the trajectories of test masses on
the motion of a binary system. The results, which are obtained
through the gradually approximate approach, can be used to search
for dark energy via the difference of the azimuth angle of the
pericenter. The classification of the motion is discussed.
\end{abstract}



\maketitle

\section{Introduction}
Several astrophysical observational data have shown that our
universe is undergoing an era of accelerated expansion
\cite{trj1,trj2,trj3}. Therefore, in order to explain this bizarre
phenomenon, various models of cosmology have been put forward,
ranging from the simplest one of a cosmological constant to the
scalar field theory of dark energy and modified gravitational theory
as well \cite{trj4,trj5,trj6}. On the other hand, taking dark energy
for the presumed cosmological component is also used, combined with
the Einstein field equations, to deal with some local gravitational
issues \cite{trj7,trj8}, such as the relevant features of black
holes. Recently, Kiselev has brought forward new static spherically
symmetric exact solutions of the Einstein equations, either for a
charged or uncharged black hole surrounded by quintessential matter
or free from it, which satisfy the condition of the additivity and
linearity \cite{trj9}. Otherwise, using a linear post-Newtonian
approach, Kerr et al. considered the orbital differential equations
for test bodies of a binary system in the Kerr-de Sitter spacetime
and gave the elliptically orbital solution \cite{trj10}. The
solutions for parabolic and hyperbolic orbits can be obtained via
the formulae of \cite{trj11}. In this paper, we study the
trajectories of test masses in a binary system under the general
metric mentioned in \cite{trj9}. Four discrete theoretical values of
the state parameter $\omega$ from $0$ to $-1$ are used to solve the
orbital equations, respectively. Similar to the classic tests of
Einstein's general relativity \cite{trj12}, the formulae gotten here
can be used to explore whether dark energy exists or not along with
the improvement upon observational techniques in astronomy. By
introducing the effective potential, we also discuss the
classification of the motion. The metric for a black hole surrounded
by quintessence is laid out in Section II. We calculate trajectories
of test masses in detail in Section III. In Section IV, we display
the classification of the motion and a brief conclusion follows. We
use the metric signature $(+,-,-,-)$ and make $G$, $\hbar$ and $c$
equal to unity.

\section{The metric for Schwarzschild space surrounded by the quintessential matter}
To begin with, Kiselev's work \cite{trj9} is reviewed briefly. The
interval of a spherically symmetric static gravitational field is
\begin{equation}
ds^{2}=e^{\nu}dt^{2}-e^{\mu}dr^{2}-r^{2}\left(d\theta^{2}+\sin^{2}\theta
d\varphi^{2}\right),
\end{equation}
where $\nu$ and $\mu$ are functions of $r$. Due to the condition of
the additivity and linearity, which implies $\mu+\nu=0$, the
energy-momentum tensor of quintessence can be written as
\begin{equation}
{T_t}^t={T_r}^r=\rho_q,
\end{equation}
\begin{equation}
{T_\theta}^\theta={T_\varphi}^\varphi=-\frac{1}{2}\,\rho_q\,(3\omega+1),
\end{equation}
where $\rho_q$ is the density of the quintessential matter while
$\omega$ is the state parameter. By setting $\mu=-\ln(1+f)$, we get
\begin{equation}
r^2f^{\prime\prime}+3(1+\omega)\,rf^\prime+(3\omega+1)f=0,
\end{equation}
whose solutions are of the forms
\begin{eqnarray}
f_{q}=\frac{\lambda}{r^{3\omega+1}}, \\
f_{\emph{BH}}=-\frac{r_\textsl{g}}{r},
\end{eqnarray}
where $\lambda$ and $r_\textsl{g}$ are the normalization factors.
Then, according to the relation
\begin{equation}
\rho_{q}=\frac{\lambda}{2}\,\frac{3\omega}{r^{3(1+\omega)}},
\end{equation}
$\lambda$ should be negative for quintessence because the density of
energy is positive. As a resulting, the metric for the space
surrounded by the quintessential matter can be expressed by
\begin{equation}
ds^{2}=\left(1-\frac{2M}{r}-\frac{\lambda}{r^{3\omega+1}}\right)dt^{2}
-\left(1-\frac{2M}{r}-\frac{\lambda}{r^{3\omega+1}}\right)^{-1}dr^{2}-r^2\left(d\theta^{2}+\sin^{2}\theta
d\varphi^{2}\right), \label{eq8}
\end{equation}
where $M$ is the black hole mass and $r_\textsl{g}=2M$. Obviously,
it can be reduced to the Schwarzschild and Schwarzschild-de Sitter
metrics by $\lambda=0$ and $\omega=-1$, respectively.

\section{Trajectory of the test mass and the motion of a binary system}
In relativistic dynamics, the contravariant 4D momentum $p^{\mu}$ is
defined as $p^{\mu}=m\frac{dx^\mu}{d\tau}$, where $m$ is the rest
mass. Then the covariant momentum, which is more important in
dynamics, can be introduced by $p_\mu=\textsl{g}_{\mu\nu}p^\nu$.
When particles move in the gravitational field, dynamical conserved
quantities are determined by the space-time symmetry of the metric
field. From the metric (\ref{eq8}), we may see clearly that $t$ and
$\varphi$ are both cyclic coordinates; therefore, conserved
quantities of the test mass are the $t$ and $\varphi$ components of
$p_\mu$, i.e.
\begin{equation}
p_0=m\left(1-\frac{2M}{r}-\frac{\lambda}{r^{3\omega+1}}\right)\frac{dt}{d\tau}=\mbox{const.},
\end{equation}
\begin{equation}
p_3=-mr^2\sin^2\theta\frac{d\varphi}{d\tau}=\mbox{const.}.
\end{equation}

Due to the symmetry, the test mass moves on the symmetric plane
formed by the initial velocity vector (3D) and center of force. We
regard the direction vertical to the orbital plane as the polar
axis. Then the motion of the test mass satisfies
\begin{equation}
\theta=\frac{\pi}{2}, \ \ \ \ \frac{d\theta}{d\tau}=0. \label{eq11}
\end{equation}
$p_0$ and $p_3$ are conserved quantities written in the form
\begin{equation}
\left(1-\frac{2M}{r}-\frac{\lambda}{r^{3\omega+1}}\right)\frac{dt}{d\tau}=E,
\end{equation}
\begin{equation}
r^2\frac{d\varphi}{d\tau}=L,
\end{equation}
where $E$ and $L$ are constants of integration that represent energy
and angular momentum per unit mass, respectively. Furthermore, the
normalization condition for the 4D velocity,
$\textsl{g}_{\mu\nu}u^\mu u^\nu=1$, also provides the equation
\begin{equation}\label{eq14}
\left(1-\frac{2M}{r}-\frac{\lambda}{r^{3\omega+1}}\right)\left(\frac{dt}{d\tau}\right)^2
-\left(1-\frac{2M}{r}-\frac{\lambda}{r^{3\omega+1}}\right)^{-1}\left(\frac{dr}{d\tau}\right)^2
-r^2\left(\frac{d\varphi}{d\tau}\right)^2=1.
\end{equation}
Equations (\ref{eq11}) to (\ref{eq14}) are four first integrals of
the geodesic equation, which compose a set of self-contained
differential equations of particle dynamics.

After  being cleared up, these equations can lead to the
relativistic extension of the Binet formula of Newtonian mechanics,
\begin{equation}
\frac{d^2u}{d\varphi^2}+u=\frac{M}{L^2}+3Mu^2+\frac{\lambda\left(3\omega+1\right)}{2L^2}u^{3\omega}
+\frac{\lambda\left(3\omega+3\right)}{2}u^{3\omega+2}, \label{eq15}
\end{equation}
where $\frac{dr}{d\varphi}\neq0$, because the orbit is not a circle
and the second term on the right-hand side is a relativistic
correction, while the last two are quintessential contributions. We
have introduced the dimensionless variable $u=GM/r$, and for
convenience $u=1/r$ has been used in the process of the calculation.

Note that $u=GM/r$ is a small quantity. When $\omega<0$, $u$ also
appears in the denominator. So we now evaluate the magnitude of the
last two terms of Eq. (\ref{eq15}). First, we make the
transformation $\alpha=-3\omega$ and $\beta=\lambda^{1/\alpha}$;
then Eq. (\ref{eq15}) turns to
\begin{equation}
\frac{d^2u}{d\varphi^2}+u=\frac{M}{L^2}+3Mu^2+\frac{1-\alpha}{2L^2}\left(\frac{\beta}{u}\right)^{\alpha}
+\frac{3-\alpha}{2}\left(\frac{\beta}{u}\right)^{\alpha}u^2.
\end{equation}
The transformation parameter $\alpha$ satisfies $\alpha \in [0,3]$,
while $\omega \in [-1,0]$. As for $\beta$, we know that the state
parameter $\omega=-1$ represents the standard $\Lambda\textrm{CDM}$
model, and therefore $\beta \sim \Lambda^{1/3}$. From recent
cosmological observations, the magnitude of the cosmological
constant is $\Lambda \approx {H_0}^2/c^2 \approx
10^{-52}\,\textrm{m}^{-2}$ with $H_0 \approx
70\,\textrm{km}\,\textrm{s}^{-1}\,\textrm{Mpc}^{-1}$. As a result,
the magnitude of $\beta$ is $\beta \sim (10^{-52})^{1/3} <
10^{-17}$. Furthermore, due to the definition $u=GM/r$, we have the
expression
\begin{equation}
\frac{\beta}{u}=\frac{\beta r}{GM}. \label{eq17}
\end{equation}
By way of a practical evaluation, we take the orbital radius of
Mercury, $r=5\times10^{10}\,\textrm{m}$, which is the smallest in
the solar system and for the other parameters we get $GM=1.5\times
10^3\,\textrm{m}$. As a result, Eq. (\ref{eq17}) yields
\begin{equation}
\frac{\beta}{u}\approx
\frac{10^{-17}\times5\times10^{-10}}{1.5\times10^3} \approx
3.3\times10^{-10} \ll u\sim10^{-7}.
\end{equation}
Furthermore, from the Binet equation in Newtonian mechanics,
\begin{equation}
\frac{d^2u}{d\varphi^2}+u=\frac{M}{L^2},
\end{equation}
whose solution is
\begin{equation}
u(\varphi)=\frac{M}{L^2}(1+e\cos\varphi),
\end{equation}
we find that $M/L^2 \sim u$. Therefore, when $\alpha>0\,(\omega<0)$
the last two terms of Eq. (\ref{eq15}) are far less than $u$ and
$M/L^2$ and can be regarded as small corrections.

According to the linear perturbation scheme put forward in
\cite{trj10}, the last three terms could be treated as perturbations
being a function of $\varphi$ by substituting for $u$ the
unperturbed solution. Due to the supernova's dimming, the dark
energy should satisfy $\omega \leq -2/3$ to fit the observations
\cite{trj13,trj14}. But here, without essential loss of generality,
we evaluate the quintessential matter from a theoretical point of
view with the state parameter in the range of $\omega \in [-1,0]$.
So we discuss the influence of quintessence on the trajectory of the
test mass in four cases.

\subsection{case of $\omega=0$}
When $\omega=0$, Eq. (\ref{eq15}) reduces to
\begin{equation}
\frac{d^2u}{d\varphi^2}+u=\frac{M+\frac{\lambda}{2}}{L^2}+\left(3M+\frac{3\lambda}{2}\right)u^2,\label{eq21}
\end{equation}
where $u$ is a small quantity, resulting in a slight correction of
$u^2$. So we ignore it in the first place, and we find the
zeroth-order approximate solution
\begin{equation}
u_0=\frac{M+\frac{\lambda}{2}}{L^2}\left(1+e\cos\varphi\right).\label{eq22}
\end{equation}
We presume that the bound orbit is an ellipse with eccentricity
$e<1$. Via substituting $u_0$ into the right-hand side of Eq.
(\ref{eq21}), we can get the first-order approximate solution
\begin{equation}
u=\left(\frac{M+\frac{\lambda}{2}}{L^2}\right)\left[1+e\cos\varphi
+3\left(M+\frac{\lambda}{2}\right)\left(\frac{M+\frac{\lambda}{2}}{L^2}\right)e\varphi\sin\varphi\right],
\label{eq23}
\end{equation}
where small higher-order quantities, which are impossible to measure
very accurately, are neglected for their insignificant effects.
Furthermore, Eq. (\ref{eq22}) means that $(M+\frac{\lambda}{2})/L^2
\sim u$, and in the weak-field regime, $M \ll r$. Introducing
$\varepsilon \equiv
3(M+\frac{\lambda}{2})[(M+\frac{\lambda}{2})/L^2]$, namely
$\varepsilon \ll 1$, makes $\cos\varepsilon\varphi \sim 1$ and
$\sin\varepsilon\varphi \sim \varepsilon\varphi$. Then Eq.
(\ref{eq23}) reduces to
\begin{equation}
u\left(\varphi\right)\approx\left(\frac{M+\frac{\lambda}{2}}{L^2}\right)\left[1+e\cos\left(1-\varepsilon\right)\varphi\right].
\end{equation}
When considering precession, we may get the difference of the
azimuth angle of the pericenter
\begin{equation}
\triangle\varphi=6\pi\left(\frac{M}{L}\right)^2+6\pi\left(\frac{\lambda
M+\lambda^2/4}{L^2}\right),
\end{equation}
where we ignore a quantity of $\varepsilon^2$, and the second term
on the right-hand side belongs to quintessence.

\subsection{case of $\omega=-1/3$}
When $\omega=-1/3$, Eq. (\ref{eq15}) reduces to
\begin{equation}
\frac{d^2u}{d\varphi^2}+\left(1-\lambda\right)u=\frac{M}{L^2}+3Mu^2.
\end{equation}
Similar to the case of $\omega=0$, the zeroth-order approximate
solution is
\begin{equation}
u_0=\frac{M}{L^2}\left[1+e\cos\left(\varphi\sqrt{1-\lambda}\right)\right];
\end{equation}
then the first-order approximate solution reads
\begin{equation}
u=\frac{M}{L^2}\left[1+e\cos\left(\varphi\sqrt{1-\lambda}\right)
+\frac{3M^2}{L^2\sqrt{1-\lambda}}e\varphi\sin\left(\varphi\sqrt{1-\lambda}\right)\right].
\label{eq28}
\end{equation}
Introducing $\varepsilon \equiv \frac{3M^2}{L^2\sqrt{1-\lambda}}$
makes Eq. (\ref{eq28}) reduce to
\begin{equation}
u\left(\varphi\right)=\frac{M}{L^2}\left\{1+e\cos\left[\left(\sqrt{1-\lambda}-\varepsilon\right)\varphi\right]\right\}.
\end{equation}
Furthermore, the difference of the azimuth angle is
\begin{equation}
\triangle\varphi=6\pi\frac{M^2}{L^2\left(1-\lambda\right)^{3/2}}+2\pi\left(\frac{1-\sqrt{1-\lambda}}{\sqrt{1-\lambda}}\right).
\end{equation}
Here, $\varepsilon^2$ is ignored, and quintessence also affects the
first term, because the zeroth-order approximate solution is changed
for the component $\varphi$.

\subsection{case of $\omega=-2/3$}
When $\omega=-2/3$, Eq. (\ref{eq15}) reduces to
\begin{equation}
\frac{d^2u}{d\varphi^2}+u=\left(\frac{M}{L^2}+\frac{\lambda}{2}\right)+3Mu^2
-\frac{\lambda}{2L^2u^2}. \label{eq31}
\end{equation}
Obviously, the form turns out to be complicated, because $u^2$
appears in the denominator. But as mentioned before, $\lambda \sim
\Lambda$, so the last two terms are both perturbation ones. Here we
could deal with it by the linear perturbation scheme in
\cite{trj10}, where the motion of test bodies in the Kerr-de Sitter
spacetime was studied. Substituting the zeroth-order approximate
solution $u_0=\frac{1}{p}\left(1+e\cos\varphi\right)$ with
$p=\left(\frac{M}{L^2}+\frac{\lambda}{2}\right)^{-1}$ for $u$ into
the r.h.s. of Eq. (\ref{eq31}), we can divide the resulting linear
equation into two parts:
\begin{equation}
\frac{d^2u_1}{d\varphi^2}+u_1=3Mp^{-2}\left(1+e\cos\varphi\right)^2,\label{eq32}
\end{equation}
\begin{equation}
\frac{d^2u_2}{d\varphi^2}+u_2=p^{-1}-\frac{\lambda
p^2}{2L^2\left(1+e\cos\varphi\right)^2}. \label{eq33}
\end{equation}
After being transformed, Eq. (\ref{eq33}) can be integrated once to
the form
\begin{equation}
\frac{d}{d\xi}\left[\left(1-\xi^2\right)^{-\frac{1}{2}}U\right] =
-\frac{q}{e\left(1-\xi^2\right)^{3/2}\left(1+e\xi\right)} +
\frac{C}{\left(1-\xi^2\right)^{3/2}},
\end{equation}
where $\xi=\cos\varphi$, $U=u_2-p^{-1}$, $q=-\frac{\lambda
p^2}{2L^2}$ and $C$ is a constant of integration. If once more we
integrate it, using formulae (2.264), (2.266) and (2.269) given in
\cite{trj15}, then we get
\begin{equation}
U=\frac{q}{1-e^2}\left[\left(1+e\Psi\sin\varphi\right)-\frac{1}{e}\cos\varphi\right]
+ C\cos\varphi+S\sin\varphi,
\end{equation}
where $S$ is a constant of integration and
\begin{equation}\label{eq36}
\Psi=\frac{1}{\sqrt{1-e^2}}\arcsin\left(\frac{e+\cos\varphi}{1+e\cos\varphi}\right).
\end{equation}
Combining with a particular solution of Eq. (\ref{eq32}), the final
solution for this case is $u=U+p^{-1}+3Mp^{-2}e\varphi\sin\varphi$,
which can be rewritten in the form
\begin{equation}
u\left(\varphi\right) \approx
\frac{q}{1-e^2}\left[\left(1+e\Psi\sin\varphi\right) -
\frac{1}{e}\cos\varphi\right]+\left(\frac{M}{L^2}+\frac{\lambda}{2}\right)
\left\{1+e\cos\left[\left(1-\varepsilon\right)\varphi\right]\right\},
\end{equation}
with $\varepsilon \equiv
3M\left(\frac{M}{L^2}+\frac{\lambda}{2}\right)$ introduced. Here $u$
may reduce to the result of Schwarzschild case \cite{trj16} as
$\lambda \rightarrow 0$.

\subsection{case of $\omega=-1$}
When $\omega=-1$, Eq. (\ref{eq15}) reduces to
\begin{equation}
\frac{d^2u}{d\varphi^2}+u=\frac{M}{L^2}+3Mu^2-\frac{\lambda}{L^2u^3}.
\end{equation}
The unperturbed solution is
$u_0=\frac{1}{p}\left(1+e\cos\varphi\right)$ with
$p=\left(\frac{M}{L^2}\right)^{-1}$. Similar to the case of
$\omega=-2/3$, according to the result (49) in \cite{trj10} the
approximate solution is
\begin{equation}
u\left(\varphi\right) \approx
\frac{q}{2\left(1-e^2\right)^2}\left[3\left(1+e\Psi\sin\varphi\right)
-\frac{1-e^2}{1+e\cos\varphi}-\left(2e+\frac{1}{e}\right)\cos\varphi\right]
+\frac{M}{L^2}\left\{1+e\cos\left[\left(1-\varepsilon\right)\varphi\right]\right\},
\end{equation}
where $q=-\lambda p^3/L^2$, $\varepsilon \equiv 3M^2/L^2$ and $\Psi$
is represented by Eq. (\ref{eq36}). When $\lambda \rightarrow 0$,
this result reduces to the Schwarzschild solution. It is different
from the two former cases in that the difference of the azimuth
angle of the pericenter hardly comes out when $\omega=-2/3$ or $-1$
due to the complicated form.

\section{Discussion and conclusion}

\begin{figure}
  \includegraphics[width=3.5 in]{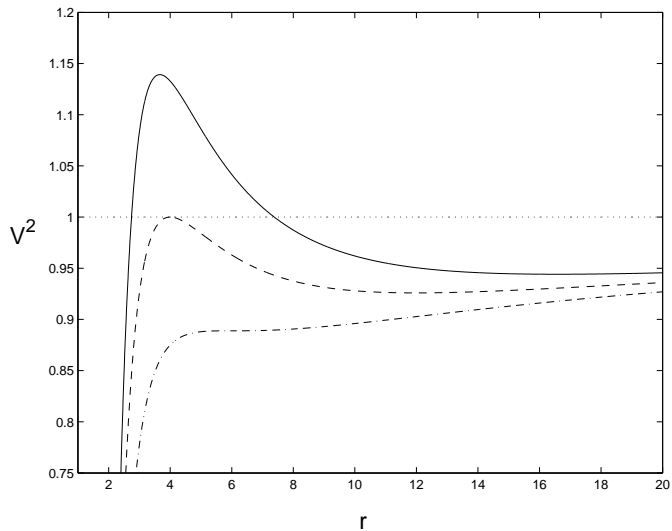}\\
  \caption{The effective potential $V^2$ for timelike geodesics for variable $L$ with $\lambda=0$ and $M=1$.
  Solid line for $L>4$, dash line for $L=4$, dash-dot line for $L=2\sqrt{3}$.}
\label{fig1}
\end{figure}

Here, we turn to a discussion of the classification of the motion.
First, we introduce the relativistic effective potential
\begin{equation}
V^2=\left(1-\frac{2M}{r}-\frac{\lambda}{r^{3\omega+1}}\right)\left(1+\frac{L^2}{r^2}\right).\label{eq40}
\end{equation}
Thus, the radial equation derived from (\ref{eq14}) can be rewritten
as $\left(\frac{dr}{d\tau}\right)^2 = E^2-V^2$, which implies that
the possible motions only occur when $E^2 \geqslant V^2$. Expand Eq.
(\ref{eq40}),
\begin{equation}
V^2 = 1 - \frac{2M}{r} + \frac{L^2}{r^2} - \frac{2ML^2}{r^3} -
\frac{\lambda}{r^{3\omega+1}} - \frac{\lambda L^2}{r^{3\omega+3}}.
\end{equation}
Because the effect of $\lambda$ on $V^2$ is very small, the curves
of $V^2$ are almost the same as that in the Schwarzschild space
\cite{trj16}. So we illustrate the effective potential $V^2$ for
variable $L$ with $\lambda=0$ in Fig. \ref{fig1}.

From Fig. \ref{fig1}, we see that for $L>4$, there is a potential
barrier whose peak value is larger than 1 when $r$ is very small,
while there is also a potential well when $r$ is large. Thus, the
possible motions can be classified as three kinds
\begin{eqnarray}
E^2 < 1, \ \ \ \ \ \ \ \ \ \ \ \ \ \ \ \ \ \ \ \text{the bound state}; \nonumber \\
1 \leqslant E^2 < V_\textrm{max}^2, \ \ \ \ \text{the scattering
state}; \nonumber \\
E^2 \geqslant V_\textrm{max}^2, \ \ \ \ \ \ \ \ \ \ \ \text{the
absorbing state}. \nonumber
\end{eqnarray}
With decreasing $L$, the central potential barrier decreases in
altitude. In the range of $2\sqrt{3} \leqslant L \leqslant 4$,
$V_\textrm{max}^2 \leqslant 1$, which indicates that the appearance
of the scattering state is impossible, with the bound and absorbing
states being left over. If $L < 2\sqrt{3}$, both the peak and the
hollow of the effective potential will disappear, and only the
absorbing state is left over. But note that the above analysis is
based on one assumption that the radius of the gravitation source is
so small that the exterior of the source is applicable for very
small $r$.

In summary, we have studied the trajectories of test bodies on the
motion of a binary system in the presence of quintessence. There are
four cases of the state parameter $\omega$ of quintessence, and we
obtain four corresponding orbital equations, where it is assumed
that the test mass moves round the center of force with eccentricity
$e<1$. The difference of the azimuth angle of the pericenter can be
cast in formulae in the two former cases, while the other two
complicated ones can hardly be done, considering their different
forms. Moreover, the effect caused by dark energy is too small to be
detectable in the solar system. But the common feature of our four
cases is that the resultant solutions reduce to the outcomes in
Schwarzschild space naturally as the parameter $\lambda$ vanishes.

\acknowledgments This work is supported by the National Natural
Science Foundation of China under Grant No. 10573004.

\end{document}